\documentclass[3p,twocolumn]{elsarticle}
\usepackage{lineno,hyperref}

\modulolinenumbers[5]

\journal{Measurement}

\usepackage{etoolbox}
\makeatletter
\patchcmd{\ps@pprintTitle}
  {Preprint submitted}
  {Authors' Accepted Manuscript submitted}
  {}{}
\makeatother

\usepackage[english]{babel}

\usepackage[T1]{fontenc}
\usepackage[latin1]{inputenc}
\usepackage{flushend}
\usepackage[]{units}
\usepackage{graphicx}
\usepackage{amssymb}
\usepackage{float}
\usepackage{amsmath}
\usepackage{mathtools}
\usepackage{hyperref}

\biboptions{sort&compress}

\usepackage[font={normalsize,it}]{caption}
\usepackage{fancyhdr}

\newcommand{\rhoff}{\rho_{\it ff}}
\newcommand{\rhoffn}[1]{\rho_{\it ff,{#1}}}
\newcommand{\rhovm}{\rho_{v}}
\newcommand{\rhovmCC}{\rho_{v}}

\newcommand{\rmi}{{\rm i}}
\renewcommand{\r}[1]{\overline{r}_{#1}}

\renewcommand{\Re}{\textrm{Re}}
\renewcommand{\Im}{\textrm{Im}}

\newcommand{\nuri}[1]{\nu_{res,#1}}
\newcommand{\nur}{\nu_{res}}
\renewcommand{\ni}[1]{\frac{\nu_{#1}}{\nu_c}}

\title{{A method based on a dual frequency resonator to estimate physical parameters of superconductors from surface impedance measurements in a magnetic field}}

\author[RM3]{Nicola Pompeo \corref{cor1}}
\ead{nicola.pompeo@uniroma3.it}

\cortext[cor1]{Corresponding author.\\
Email: nicola.pompeo@uniroma3.it.\\
\url{https://doi.org/10.1016/j.measurement.2021.109937}.\\
Available online 29 July 2021.\\
Distributed under CC-BY-NC-ND license.
}

\author[RM3]{Kostiantyn Torokhtii}
\ead{kostyantin.torokhtii@uniroma3.it}

\author[RM3]{Andrea Alimenti}
\ead{andrea.alimenti@uniroma3.it} 

\author[RM3]{{Enrico Silva}}
\ead{enrico.silva@uniroma3.it}

\address[RM3]{Dipartimento di Ingegneria, Universit\`a Roma Tre, Via Vito Volterra 62, 00146 Rome, Italy}

\begin{document}
\begin{abstract}
{
High frequency applications of superconductors in a dc magnetic field rely on the estimate of the characteristic crossover frequency $\nu_c$ between low and high losses. Customarily, high sensitivity resonant techniques, intrinsically operating at discrete frequencies, are used to estimate $\nu_c$. We exploit here a method based on a dual frequency resonator. We show that single-frequency evaluations of $\nu_c$ lead to heavy underestimations of the superconductor surface resistance. We describe a combined analytical and experimental approach that gives more accurate estimates for $\nu_c$, and additionally allows to test the underlying physical model.
}
\end{abstract}


\maketitle

\section{Introduction}
\label{sec:intro}
{Cutting edge fundamental research fields relies heavily on}
the low losses of superconductors at {radio and microwave} frequencies. As a 
{first } example, particle accelerators require superconducting cavities \cite{Posen2017,Cichalewski2020} in which the extremely low losses allow for high cavity quality factor $Q$, and then strong accelerating electric fields of the order of tens of MV/m. 
The power handling capabilities of the next generation particle accelerators will be so demanding that also the beam screens, necessary to confine the {synchrotron} radiation emitted by the charged, high-energy particles over their {bent } trajectories are thought to require superconductors: the entire length of the 100-km long beam tunnel of the Future Circular Collider {(FCC) }could be covered by a suitable superconductor. As a striking difference with respect to well-established applications of high-frequency superconductivity, the superconductors should keep the losses much lower than those of copper {at frequencies up to $\sim$2~GHz } in very intense static magnetic fields at the scale of 16~T \cite{Calatroni2017b, Abada2019}.
A similar requirement -- superconductors operating at microwave frequencies in a dc field of several tesla -- is shared by dark matter hunting experiments: axions, possible candidates for the elusive dark matter, are predicted to interact with the electromagnetic field of high $Q$ resonators, tuned to their mass frequency equivalent, within static magnetic fields, so that setups with superconducting cavities have been designed and tested \cite{Alesini2019,Alesini2020}. The sensitivity of the instrument is proportional to $Q\cdot B^2$, where $B$ is the magnetic flux density, so it is clear that low losses and high dc fields are requested.
{The physical quantity determining the electromagnetic response of a conductor (and of a superconductor) in the high frequency regime is the surface impedance $Z$. This is a complex quantity, defined as $Z\coloneqq E_{t}/H_{t}$, ratio of the tangential components of the electric $E_{t}$ and magnetic $H_{t}$ field on the superconductor surface \cite[\S 2.6]{Collin1992}\cite[\S 87]{Landau1981}\cite[\S 8.1]{jackson1999book}. }
Hence, the prediction, determination and optimization of the surface impedance $Z$ of superconductors, and in particular of the surface resistance {$R\coloneqq\Re(Z)$ } which governs the power losses {\cite[\S 2.9, \S 7.4]{Collin1992}\cite[\S 8.1]{jackson1999book}}, are of renewed and paramount relevance.
The expression $Z=\sqrt{\rmi2\pi\nu\mu_0\rho}$ connects the complex resistivity (material property) $\rho$ to $Z$ \cite{Collin1992} at the frequency $\nu$ ($\mu_0$ is the magnetic permeability) in (super)conductors with thickness $d$ larger than the normal $\delta$ and London $\lambda$ penetration depths \cite{PompeoImeko2017a}.
In a static magnetic field, however, the losses in the ubiquitously used type II superconductors are dominated by far by the motion of quantized flux tubes ("fluxons" or "vortices") \cite{Tinkham1996book}. Fluxons oscillate under the Lorentz force exerted by the high frequency currents, giving rise to electrical fields with {components } parallel to the currents, thus yielding power dissipation. Their motion is hindered by material defects that "pin" the vortices, exerting an elastic recall. The beneficial (from the point of view of limited power dissipation) effect of pins can be weakened by thermal activating depinning (often referred to as fluxon creep).

The widely accepted model \cite{Tinkham1996book} describes fluxons under a high frequency current as damped harmonic oscillators, with an additional stochastic thermal force \cite{Golosovsky1996, PompeoLTP2020}. The resulting expression for the  complex vortex motion resistivity $\rho_{v}$ can be written down as \cite{Pompeo2008}:
\begin{eqnarray}
\label{eq:rhovmCC}
\rhovmCC=\rhoff\frac{\chi+\rmi\frac{\nu}{\nu_c}}{1+\rmi\frac{\nu}{\nu_c}}=\rho_{v1}+\rmi\rho_{v2}
\end{eqnarray}
where $\rhoff\propto B$ \cite{Tinkham1996book} is the so-called flux flow resistivity, corresponding {to } the fluxon motion free from pinning and thermal activation{; }  
$\chi\in[0,1]$ is an adimensional creep factor, mapping absent to total thermal activation; $\nu_c$ is a characteristic frequency, depending on $\chi$ and on the so-called (de)pinning frequency $\nu_p$ through a function which depends on the specific model used \cite{Pompeo2008}. The latter equality defines the real and imaginary parts of $\rho_v$. At zero creep, $\nu_c\rightarrow\nu_p$.

\begin{figure}[hbt!]
\centering
\includegraphics[width=0.8\columnwidth]{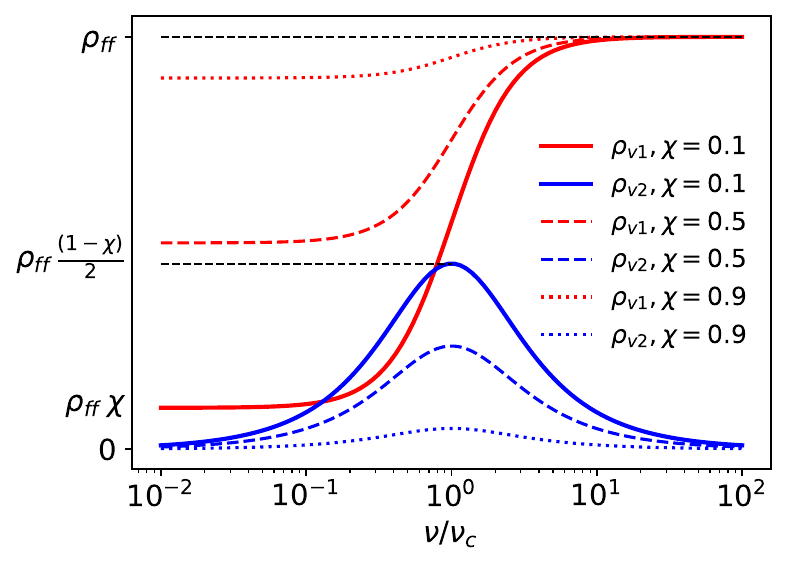}
\caption{Frequency dependence of the real and imaginary parts of $\rhovmCC$, Eq. \eqref{eq:rhovmCC}, plotted for $\chi=\{0.1, 0.5, 0.9\}$.}
\label{fig:rhovmCCvsNu}
\end{figure}
Figure \ref{fig:rhovmCCvsNu} reports $\rho_{v}$, Eq. \eqref{eq:rhovmCC}, at a fixed field and temperature and as a function of the frequency. The role of the various parameters is graphically depicted. $\rhoff$ and $\rhoff\chi$ set the upper limit and the lower limit of the real resistivity, respectively, and the characteristic frequency $\nu_c$ marks the crossover between the low-{frequency}, low dissipation from the high-frequency, high dissipation regime. As such, $\nu_c$ or the zero-creep limit $\nu_p$ are often used to synthetically describe the performances of a superconductor for high frequency applications. Remarkably, very little studies exist in very high dc field microwave superconductivity, and the applied methods usually rely on older models \cite{Gittleman1966} with $\chi=0$. 
The latter is a questionable choice for high-$T_c$ superconductors: 
their high operating temperature ($T_c\sim30-100\;$K) and their layered microscopic structure, make the thermal effects much more effective than in low-$T_c$ ($T_c\sim1-10\;$K), isotropic superconductors. Moreover, when accurate measurements are performed, even in the latter the thermal effects can be detected \cite{Song2009, Silva2011}.
This paper describes a method to accurately measure all the three relevant vortex parameters.

Wideband {(i.e. over more than a decade in frequency)}, swept-frequency measurements are the first obvious choice to test Eq. \eqref{eq:rhovmCC} and to derive the vortex parameters. Although the cryomagnetic environment severely limits the accuracy of such techniques \cite{Silva2016}, a few attempts were successful in particular with the so-called Corbino disk technique \cite{Silva2016, Booth1994, Kitano2008a, Felger2013}. 

Low sensitivity and difficult calibration (in a cryogenic environment) relegate the wideband techniques to niche studies, in favour of the more sensitive and widely used resonator-based methods 
\cite{Kobayashi1998, Jacob2001, Lue2002, Cherpak2003, Hanaguri2003, Hashimoto2003, Cherpak2005, Ghigo2005a, Huttema2006, Yeh2013, Pompeo2014, Ohshima2018, Alimenti2019a}.
{The latter have been also codified as an International Standard \cite{iec} for what concerns the high sensitivity measurement of $R$ in zero magnetic field only}.

Resonators work intrinsically at fixed frequencies dictated by the resonant mode of operation. 
Two observables can be measured, the real and imaginary part of $\rho_{v}$, against three parameters. The common approach neglects $\chi$ altogether. Although estimates for $\rhoff$ and $\nu_p$ can still be given, it can be shown that the associated uncertainties can reach and exceed 100\% \cite{Pompeo2008}. 

Multi-mode resonators, capable of operating at multiple discrete frequencies, can be a viable alternative \cite{Kobayashi1998, Powell1998}. In superconductors, planar resonators are typically exploited, with the need for thin film, patterned in the needed geometry.

Here we propose the exploitation of a purpose-built dual mode Hakki-Coleman \cite{Hakki1960} dielectric loaded resonator, operating on two quasi-{transverse electric (TE) } modes with spaced resonant frequencies $\sim16\;$GHz and \mbox{$\sim27\;$GHz}. Surface perturbation method is employed, to allow for the measurement of pristine samples.
The aim is the determination of {all the main } vortex parameters, in particular the characteristic frequency $\nu_c$ red{but including also the often neglected creep factor}, of superconducting thin films. {A suitable method for their extraction from dual frequency measurement is presented, with an emphasis on the uncertainty evaluation and their minimization by suitably selecting the measuring frequencies. }
Preliminary results were reported in Ref. \cite{PompeoIMEKO2020}.

The paper is organized as follows. In Sec.~II the measurement technique is briefly described, {providing {the }necessary background. In Sec. \ref{sec:method} we discuss the data analysis method, including the optimization of the uncertainties on the  vortex parameters derived by dual frequency measurements.}
Sample measurements and results are reported in Sec. \ref{sec:experimental}. Short conclusions are drawn in Sec. \ref{sec:conclusions}. 

\section{The measurement technique}
\label{sec:technique}
\begin{figure}[htbp]
\centering
\includegraphics[width=0.8\columnwidth]{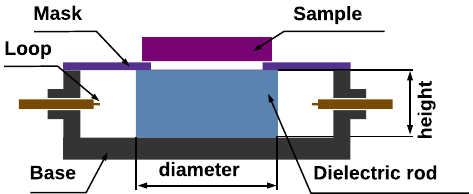}
\caption{Sketch of the dielectric resonator. {The coupling is made through loops laying on the plane normal to the section plane here depicted, in order to couple with the magnetic field lines.}}
\label{fig:sketch}
\end{figure}

Surface impedance measurements are performed by means of a copper cylindrical electromagnetic resonator, 
loaded with a single crystal sapphire rod (diameter 7.13(1) mm, height 4.50(1) mm), as reported in Fig. \ref{fig:sketch}. 
{The dielectic-loaded resonator (DR) design process has been a standard one: we have built up a mode chart by means of finite-element electromagnetic numerical simulations with commercially available software, we have optimized the dielectric rod size to obtain (i) good frequency separation of the desired modes from the unwanted ones, (ii) high $Q$, (iii) high sensitivity for samples of typical size $10\times10\;$mm$^2$, with the constraints given by our measurement apparatus, in which the cryostat has to be placed inside a magnet. For additional information about the design process and accuracy matters we refer to specialized papers \cite{Alimenti2019a, Pompeo2017a}, together with Ref. \cite{Hashimoto2003} as an excellent example about standard steps in the design. }
The resonator is used in the end-wall replacement configuration with samples of size $\sim1\;$cm$^2$ placed on a base and covered by a thin metal mask, with a central hole of diameter  6.50(1) mm in order to preserve the circular symmetry of the structure. 
The resonator operates on the quasi-transverse electric TE$_{0l1}$ ($l=1,2$) modes, with resonant frequencies $\nuri1=16.4\;$GHz and $\nuri2=26.6\;$GHz.
{
The operating frequencies are chosen in view of the applications and the reasonably easy interpretation of the data. Applications, such as the mentioned FCC beam screen, require a characteristic frequency in the range of tens of GHz. Thus, it is appropriate to experimentally probe a range close to the desired values in order to have maximum sensitivity. Second, the frequency needs to be kept sufficiently high so that the fluxon motion is limited to tiny oscillations, a requirement for Eq. \ref{eq:rhovmCC} whence the fluxon parameters.}

The external microwave line is coupled to the resonator by means of loop antennas. The resonator is operated in transmission, and {the two-ports } scattering coefficients $S_{ij}$ ($i,j=1,2$) are measured by a Vector Network Analyzer, with frequency sweeps around each of the resonant frequencies.

The cryogenic environment {prevents }a full calibration. The uncalibrated sections of the transmission lines and other non-idealities were taken into account by the fit procedure \cite{Pompeo2017a, Torokhtii2019a}, in order to measure the quality factor $Q$ of the resonator and the resonant frequency $\nur$ for each mode. 

{Since from now on we focus on magnetic field $H$ induced variations, in the following, for a generic physical quantity $A$, $A(H)$ denotes its value for the value $H$ of the magnetic field, $A(0)=A(H=0)$ and:
\begin{eqnarray}
\label{eq:delta}
\Delta A(H)\coloneqq A(H)-A(0)
\end{eqnarray}
}
To derive the vortex parameters, the field variation of the surface impedance {{$\Delta Z(H)=\Delta R(H)+\rmi \Delta X(H)$ --~where $X\coloneqq\Im(Z)$ denotes the surface reactance~-- } at fixed temperature $T$ is needed. Within the small perturbation approach,
{the superconducting sample contributes to the DR $Q$ and resonant frequency shift $\Delta \nur$ through \cite{Staelin1994book}:
\begin{subequations}
\label{eq:Z}
\begin{align}
R(H)&=\frac{G}{Q(H)}-{background_R}\\
\Delta X(H)&=-2G\frac{\nur(H)-\nur(0)}{\nur(0)}+{background_X}
\end{align}
\end{subequations}
}
where $G$ is a mode-dependent geometrical factor, determined through {finite elements electromagnetic } simulations,
{and $background_R$ and $background_X$ are the DR contributions (due the its bases, the mask, the lateral wall and the dielectric rod) to $Q$ and $\nur$, respectively. Since the DR is made of non-magnetic materials, these quantities yield no/negligible variations by varying the applied magnetic field. Hence, focusing on $\Delta Z(H)$:  }
\begin{subequations}
\label{eq:DZ}
\begin{align}
\Delta R(H)&=G\left(\frac{1}{Q(H)}-\frac{1}{Q(0)}\right)\eqqcolon G\Delta r(H)\\
\Delta X(H)&=-2G\frac{\nur(H)-\nur(0)}{\nur(0)}\eqqcolon G\Delta x(H)
\end{align}
\end{subequations}
where $\Delta r$ and $\Delta x$ {are defined through the last equalities in both equations and represent normalized } measured quantities, useful for the discussion in Sec. \ref{sec:method}.

Examples of actual resonance curves as measured in terms of the scattering coefficient $|S_{21}|$ for both modes are reported in Fig. \ref{fig:resCurves}. 
\begin{figure}[hbt!]
\centering
\includegraphics[width=0.8\columnwidth]{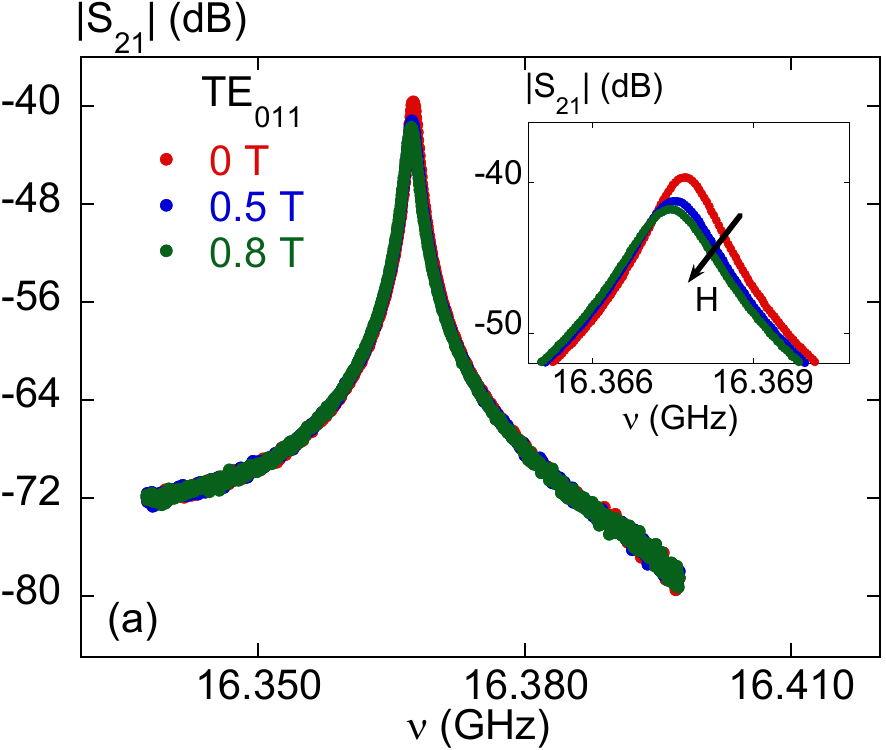}
\includegraphics[width=0.8\columnwidth]{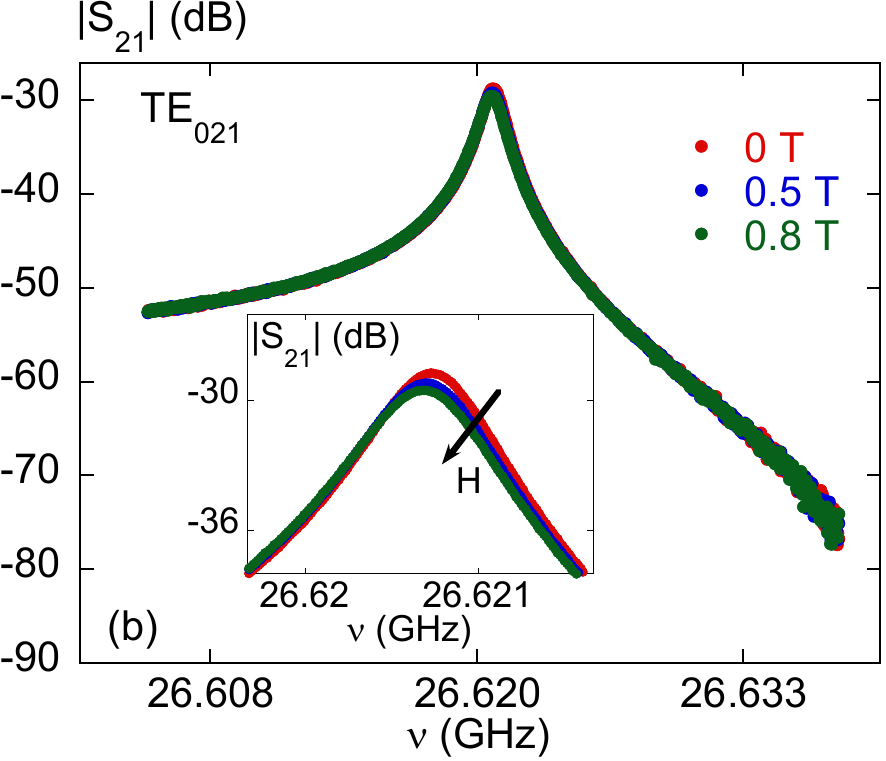}
\caption{Modulus of the scattering coefficient $S_{21}$ of the resonator vs $\nu$ around the resonant frequencies of the two modes (TE$_{011}$ in panel (a), TE$_{021}$ in panel (b)), measured at fixed $T=10.00(5)\;$K and at elected magnetic fields $H$. The change in the resonator response due to the sample $Z(H)$ is apparent. In the insets, zooms of the curves.}
\label{fig:resCurves}
\end{figure}
The resonator is loaded with the superconducting sample described in Sec. \ref{sec:experimental}. Measurements are taken at the fixed temperature ${T=10.00(5)\;{\rm K}}$ and at selected applied magnetic field intensities. The changes of $Q$ and $\nur$ with the applied field reflect the changes in $Z$ due to the motion of flux lines, according to Eq. \eqref{eq:DZ}. 
{The field dependent $Q(H)$ and $\nur(H)$ for the two modes are reported in Fig. \ref{fig:DRH}, where is apparent the magnitude of the superconductor surface impedance field induced variations with respect the absolute values of $Q$ and $\nur$.}
\begin{figure}[hbt!]
\centering
\includegraphics[width=0.8\columnwidth]{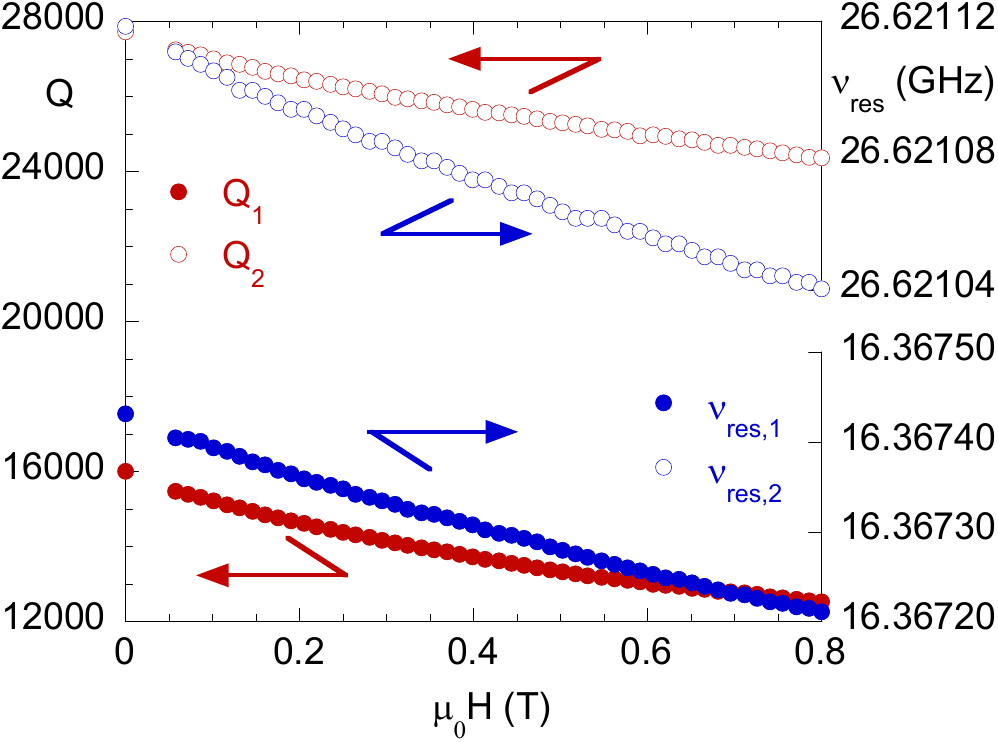}
\caption{{Field dependent $Q_i(H)$ and $\nuri{i}(H)$, with $i=\{1, 2\}$ for TE$_{011}$ and TE$_{021}$ resonant mode, respectively, measured at fixed $T=10.00(5)\;$K. The change in the resonator response due to the sample $Z(H)$ is apparent.}}
\label{fig:DRH}
\end{figure}
{Finally, the obtained resonant frequencies are well in agreement with those computed through finite element electromagnetic simulations, thus allowing a clear identification. }
In films of thickness ${d\ll \min(\delta, \lambda)}$
the so-called thin-film approximation {applies } \cite{PompeoImeko2017a}:
\begin{eqnarray}
\label{eq:rhovmexp}
\Delta Z(H)\simeq\frac{\rho_{v}(B)}{d}
\end{eqnarray}
with $B\simeq\mu_0H$ in the London limit.

\section{Method}
\label{sec:method}
{
\subsection{General aspects}
\label{sec:general}
{
The complex vortex motion resistivity $\rho_{v}$ is obtained at the frequencies $\nuri{i}$, 
{(with $i=\{1, 2\}$
referring to the first TE$_{011}$ and second (TE$_{021}$) resonant mode, respectively)
by }
combining measurements of {$Q_i$ } and $\nuri{i}$ at fixed $T$ and variable $H$ (Eq. \eqref{eq:DZ}), with the thin film approximation Eq. \eqref{eq:rhovmexp}, to get:
}
\begin{eqnarray}
\label{eq:sys1}
\nonumber
\rho_{v,i}&=&\rho_{v1,i}+\rmi\rho_{v2,i}=\\ \nonumber
&=&G_i d\left[\Delta \frac{1}{Q_i}+\rmi\left(-2\frac{\Delta\nuri{i}}{\nuri{i}}\right)\right]=\\
&=&G_i d(\Delta r_i + \rmi\Delta x_i)
\end{eqnarray}
where $\Delta$ denotes the field variations as in {Eq. \eqref{eq:delta}, {$\rho_{v1}\coloneqq{\rm Re}(\rho_v)$} and ${\rho_{v2}\coloneqq{\rm Im}(\rho_v)}$}.
In Eq. \eqref{eq:sys1} only experimental quantities are present.
{Once $\rho_v$ is obtained from measurements, vortex parameters are obtained within the model of Eq. \eqref{eq:rhovmCC}.
An important point is the evaluation of the uncertainties involved. In this paper, standard deviations are used to evaluate the uncertainties (coverage factor $k=1$).

Before addressing the possible methods to obtain the vortex parameters, in particular $\nu_c$, it is useful to recall some features of the theoretical model{.}

{Fig. \ref{fig:rhovmCCvsNu} reports a plot of the real and imaginary parts of $\rhovmCC$ vs $\nu$ (note the logarithmic scale for the frequency) for selected values of $\chi=\{0.1, 0.5, 0.9\}$.}
The real part increases monotonously, from a low frequency value $\rhoff
\chi$ to the asymptotic high frequency limit $\rhoff$; the imaginary part is vanishingly small in the above two limits (hence, from a measurement point of view this implies a small measurable signal) and attains a maximum value $\rhoff(1-\chi)/2$ at $\nu_c$. 
By increasing the creep factor $\chi$ from 0 to 1 (physically, with increased thermal processes), the above features wash out, ultimately yielding constant {$\rho_{v1}\rightarrow\rhoff$ and $\rho_{v2}\rightarrow 0$ } and preventing a precise determination of $\nu_c$. 
Since the focus is on the accurate determination of $\nu_c$, one should carefully choose the measuring frequencies $\nu_i$, {given by the resonant frequencies $\nuri{i}$ of the DR. }
By focusing the attention on the real part only, one would expect a maximum sensitivity to $\nu_c$ {through } measurements on the maximum slope of $\rho_{v1}$, with a choice $\nu_{1} < \nu_c < \nu_2$, and $\nu_i$ sufficiently separated and not too far from $\nu_c$. It will be clear later that this is not the only good choice from the point of view of reduced uncertainties.

We describe now two main approaches to the determination of the vortex parameters: standard multiple curve fitting and a closed-form, analytical solution by inversion of Eq. \eqref{eq:sys1}.
We show below that the closed-form approach leads, in some circumstances, to lower uncertainties than {a }plain fitting. Moreover, it can be used to optimize the technique by determining the best choice of the measurement frequencies to minimize the uncertainty on the most important quantity, $\nu_c$.\footnote{It is worth noting that the ratio between the measurement frequencies cannot be completely freely chosen, because of the constraints related to the actual resonators and modes that can be used, so that an indication of a range of acceptable values for $\nu_i$, around the optimal values, is necessary.} Finally, a combination of the methods can help to shed light on the consistency of the model with the experimental data.

\subsection{Standard curve fitting}
The standard curve fitting, based on the non-linear least squares Levenberg-Marquardt algorithm \cite{Zic2016, More1980},
fits {at each $H,T$ the four experimental points $\rho_{v1}(\nu_1), \rho_{v2}(\nu_1), \rho_{v1}(\nu_2), \rho_{v2}(\nu_2)$ 
to the theoretical Eq. \eqref{eq:rhovmCC}.} 
Its output{s } are the fit parameters ($\nu_c$, $\chi$ and $\rhoff$) with their uncertainties, determined from the covariance matrix of the fit parameters.  
We refer to this approach as ``agnostic'': it is applied without exploiting any particular mathematical properties of Eq. \eqref{eq:rhovmCC} or physical properties of the underlying model. Moreover, the adherence of the data to the model is neither questioned nor checked.
{Indeed, inconsistencies of the data with the theoretical model do not emerge from the fit procedure, that provides the fitting parameters anyway, possibly with increased uncertainties. We further comment on this point in Sec. \ref{sec:rhoff}.}

\subsection{Closed form inversion: $\nu_c$ and $\chi$}

The analytical solution in closed form can be computed from the overdetermined system of four equations obtained {by}  equating the measurements at the two frequencies of {the real and imaginary parts of} $\rho_{v}$, four observables, to Eq.~\eqref{eq:rhovmCC}, containing the three unknowns ($\nu_c$, $\chi$ and $\rhoff$):
\begin{subequations}
\label{eq:system}
\begin{align}
\rhoff\frac{\chi+\left(\frac{\nu_1}{\nu_c}\right)^2}{1+\left(\frac{\nu_1}{\nu_c}\right)^2}=G_1 d \Delta r_1\\
\rhoff\frac{(1-\chi)\frac{\nu_1}{\nu_c}}{1+\left(\frac{\nu_1}{\nu_c}\right)^2}=G_1 d \Delta x_1\\
\rhoff\frac{\chi+\left(\frac{\nu_2}{\nu_c}\right)^2}{1+\left(\frac{\nu_2}{\nu_c}\right)^2}=G_2 d \Delta r_2\\
\rhoff\frac{(1-\chi)\frac{\nu_2}{\nu_c}}{1+\left(\frac{\nu_2}{\nu_c}\right)^2}=G_2 d \Delta x_2
\end{align}
\end{subequations}
{In the above, the subscripts $i=\{1, 2\}$ in the quantities $\Delta r_i$, $\Delta x_i$, $G_i$, similarly to what already done for $\nu_i$, indicate the first (TE$_{011}$) and second (TE$_{021}$) resonant mode, respectively.
}
Here the measuring frequencies $\nu_i$ correspond to $\nuri{i}$ in the experiment. In the following computation of the uncertainties on the vortex parameters, $u(\nu_i)$ 
{is the main source of uncertainty on $\Delta X$, but its other contributions to the overall uncertainties on the vortex parameters can be neglected\footnote{{Indeed, considering that $\nu_i$ is normalized to $\nu_c$, the typical value $u(\nu_i)/\nu_i\sim 10^{-7}$ is negligible with respect to $u(\nu_c)/\nu_c\sim 10^{-2}-10^{-1}$ which is obtained by taking $u(\nu_i)=0$ (see later).}}.}
In order to solve Eq.s \eqref{eq:system}, it proves helpful to define the following ratio quantities:
\begin{subequations}
\label{eq:r}
\begin{align}
\r1&\coloneqq\frac{\rho_{v2,1}}{\rho_{v1,1}}=\frac{\Delta x_1}{\Delta r_1}=\frac{\ni1(1-\chi)}{\chi+\left(\ni1\right)^2}\\
\r2&\coloneqq\frac{\rho_{v2,2}}{\rho_{v1,2}}=\frac{\Delta x_2}{\Delta r_2}=\frac{\ni2(1-\chi)}{\chi+\left(\ni2\right)^2}
\end{align}
\end{subequations}
We stress that the second equality expresses the ratios $\r i$ in terms of purely experimental quantities
(the geometrical factors $G_i$ and the thickness $d$ cancel out), while the last equality makes the connection with the theoretical expression, Eq. \eqref{eq:rhovmCC}, and only two unknowns, $\nu_c$ and $\chi$, appear. Thus, Eq.s\eqref{eq:r} can be solved providing analytical expressions for $\nu_c(\r1, \r2)$ and $\chi(\nu_c, \nu_i)$.
Explicit expressions are reported and discussed in \ref{sec:appendix}. Analytical constraints allow to establish the range of admissible values of $\r i$ as:
\begin{eqnarray}
\label{eq:domain}
\frac{\r1}{p}\leq\r2\leq p\r1 
\end{eqnarray}
where {$p\coloneqq\nu_2/\nu_1>1$}. This is an important analytical outcome, since it provides a simple experimental test of the applicability of the model, Eq. \eqref{eq:rhovmCC}, to the data set. Moreover, on more practical grounds, it allows also to check if the system can be actually solved: in the case of
data close to the domain boundaries, the random uncertainties on $\r{i}$ may push the data outside of the allowed domain,} {preventing the computation of $\nu_c$.}
{Two } additional conditions are related to the scale factors. The first is trivial{: } $\rho_{v1,1}<\rho_{v1,2}$ (see Fig. \ref{fig:rhovmCCvsNu}). The {second }refers directly to the determination of $\rhoff$, that we address in the next subsection.

\subsection{Closed form inversion: determination of $\rhoff$}
\label{sec:rhoff}
{Once $\chi$ and $\nu_c$ are determined by Eq.s \eqref{eq:r}, $\rhoff$ can be obtained from the remaining two equations from \eqref{eq:system}. Explicit expressions are given in Sec. \ref{sec:appendix}.
Since the problem is overdetermined (two remaining equations, one unknown), the two solutions $\rhoffn1$ and $\rhoffn2$ (that should be equal within the uncertainties) can be compared as a consistency check.
If $\rhoffn1$ and $\rhoffn2$ overlap within the} {uncertainties, the experimental data is fully compatible with the theoretical model, and $\rhoff$ can be taken as their average. }
Otherwise, either (i) the values of $G_i$ are not correct and/or some uncertainties have been underestimated, thus indicating a revision of the data analysis, or (ii) the experimental data are not compatible with the theoretical prediction of Eq. \eqref{eq:rhovmCC}. {The latter } is a result \textit{``per se''} 
{since the model can and is customarily (e.g. in design processes) used to estimate the expected surface impedance of a superconductor at frequencies different from the measurement ones. It is clear that if in a given superconducting material the model cannot be applied, any extrapolation based on it would not be reliable.}

By contrast, the ``agnostic'' fit-based approach is certainly more robust, apart from possibly larger uncertainties, in that it gives a result anyway, but it cannot provide this important check.

\subsection{{Optimal choice of the measuring frequencies}}
\label{sec:optimal}
From the analytical solutions for $\nu_c$ and $\chi$ the uncertainties $u(\nu_c)$ and $u(\chi)$ can be easily computed in closed form (see \ref{sec:appendix}). This feature is helpful in finding the optimum pair of measuring frequencies $\nu_i$ which minimizes $u(\nu_c)$.
This can be useful both in the design process of a resonator, and to determine which uncertainty to expect with a given resonator.
The numerical study takes advantage from working in normalized frequencies, $\nu_1/\nu_c$ and $\nu_2/\nu_c$. To reduce the parameter space {to be explored}, we take ${u(\Delta r_i)= u(\Delta x_i)}$, as reasonable in typical real setups. The resulting uncertainty depends also on the combined parameter {$s\coloneqq G_2 u(\Delta r_2)/G_1 u(\Delta r_1)$ } (see \ref{sec:appendix}, Eq. \eqref{eq:sensitivity_nu0}).

We thus explore the parameter space ${\nu_1/\nu_c < \nu_2/\nu_c}$ in the range $[0,5]$ for selected values of $\chi$ and $s$.\footnote{For the resonator described in Sec. \ref{sec:experimental}, we have $s\simeq 2$. The overall scale factor $u(\Delta r_1)G_1 d /\rhoff$, see Eq. \eqref{eq:sensitivity_nu0}, is computed for the resonator and the measurement shown in Sec. \ref{sec:experimental}.} 
Some results for sample values $\chi$ and $s=2$ are reported in Fig.s \ref{fig:sensitivity}.
\begin{figure}[hbt!]
\centering
\includegraphics[width=0.8\columnwidth]{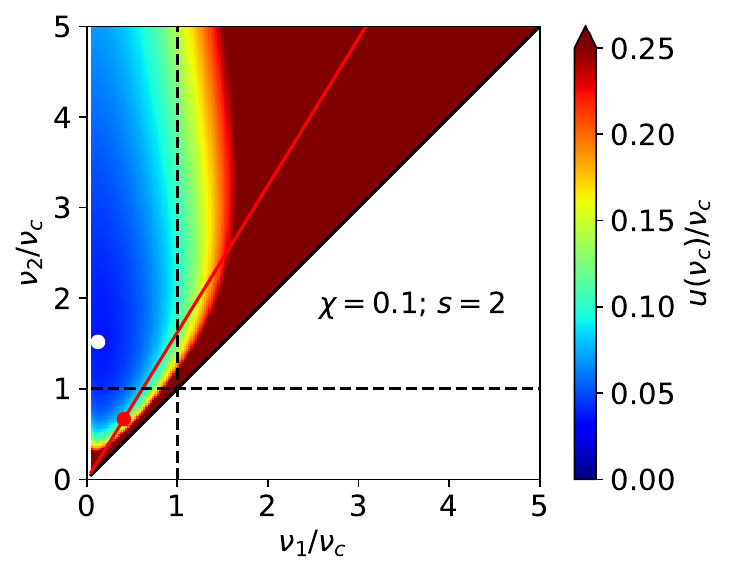}
\includegraphics[width=0.8\columnwidth]{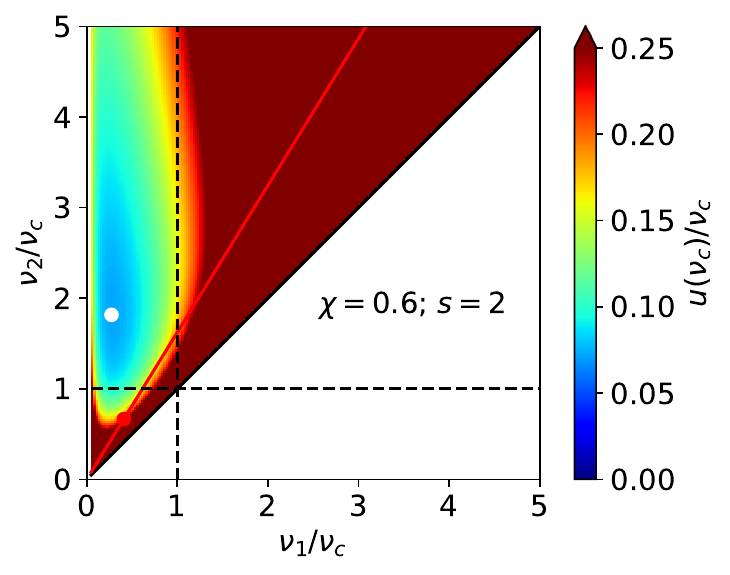}
\caption{Relative uncertainty $u(\nu_c)/\nu_c$ vs $(\nu_1, \nu_2)$ computed for $s=2$ (value for the resonator described in Sec. \ref{sec:experimental}) and selected values of $\chi$ ($0.1$ and $0.6$, upper and lower panel respectively). 
White area: $\nu_1/\nu_c > \nu_2/\nu_c$, values not computed. }
\label{fig:sensitivity}
\end{figure}
\begin{figure}[hbt!]
\centering
\includegraphics[width=0.8\columnwidth]{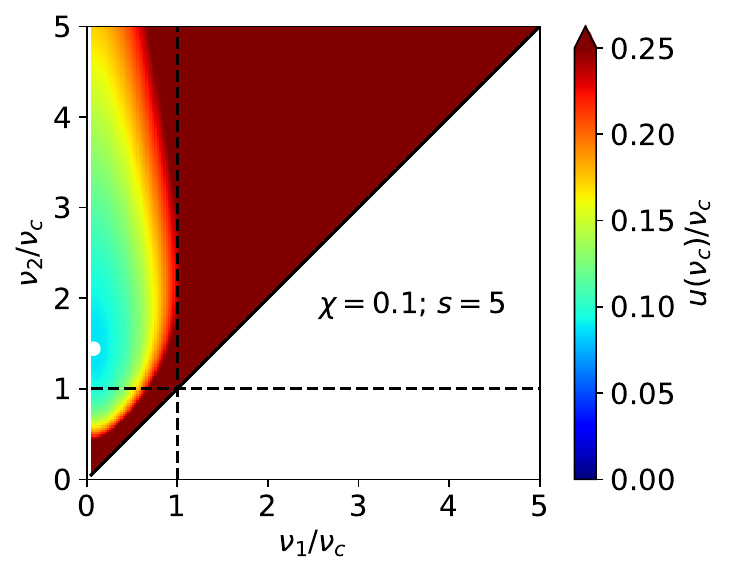}
\includegraphics[width=0.8\columnwidth]{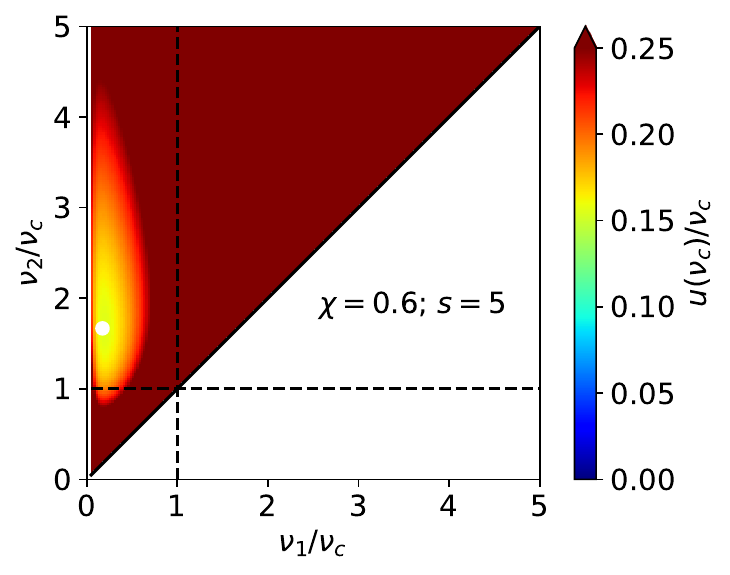}
\caption{Relative uncertainty $u(\nu_c)/\nu_c$ vs $(\nu_1, \nu_2)$ computed for $s=5$ and selected values of $\chi$ ($0.1$ and $0.6$, upper and lower panel respectively).}
\label{fig:sensitivity_worst}
\end{figure}
The horizontal and vertical dashed lines correspond to ${\nu_2=\nu_c}$ and ${\nu_1=\nu_c}$, respectively.
It can be seen that the smaller uncertainties (blue region) are obtained mainly when $\nu_1<\nu_c$ and $\nu_2>\nu_c$, as anticipated in Sec. \ref{sec:general}.
An important result is that also $\nu_1<\nu_c$ and $\nu_2<\nu_c$ can lead to acceptable uncertainties, depending on the values of $\chi$ (higher $\chi$ imposes more stringent choices on $\nu_i$).
The normalized pair $(\nu_1, \nu_2)$ corresponding to the absolute minimum uncertainty {$u(\nu_c)$ } is marked as a white dot, well within the blue region.

The main features are not affected by the values of $s$ and $\chi$: as an example, in Fig. \ref{fig:sensitivity_worst} a case with a worst, higher $s=5$ is presented.

The red continuous lines {Fig. \ref{fig:sensitivity} represent the working region of the dielectric resonator presented in Sec. \ref{sec:experimental}, where $p=1.62$. The intersection of this line with yellow region, corresponding to a maximum accepted relative uncertainty ${u(\nu_c)/\nu_c\sim15\;\%}$, yields the range of possible $\nu_c$ (in this case 10--60 GHz) which can be determined within the asked uncertainty. 
The actual working point on this line depends on the $\nu_c$ value of an actual sample: for example, the red point marks the position for the measurement described in the next Section, which yields ${\nu_c=40\;{\rm GHz}}$ for $\mu_0 H=0.5\;$T.}
\subsection{{Comparison between closed form inversion and fitting approaches}}

The system of equations \eqref{eq:r} does not depend on $G_i$ and $d$. Thus, in the inversion method $u(\nu_c)$ and $u(\chi)$ are not affected by the uncertainties on these two quantities, and only the evaluation of $\rhoff$ depends on the products $G_i d$.

The ``agnostic'' fit-based approach, on the other hand, does not exploit the normalized $\r{i}$ and thus is affected by the uncertainties on $G_i$ and $d$. 
The thickness $d$ is a common scale factor for all the $\Delta x_i$ and $\Delta r_i$: it does not modify the observed frequency dependence and thus does not impact on $\nu_c$ and $\chi$.

The two geometrical factors, however, do not behave as a common scale factor: they affect in different fashions the measured quantities at the different frequencies, yielding an increased uncertainty on $\nu_c$ and $\chi$ with respect to the analytical inversion approach. Additionally, since they} {are related to modes sustained by the same resonator,} {they are} {affected by the same uncertainties on the dimensions of the resonator and on the permittivity of the dielectric,} {so that $G_1$ and $G_2$ have correlated uncertainties}.
Obviously, these effects are visible if the uncertainties on $G_i$ are not negligible with respect to those on $\Delta r_i$ and  $\Delta x_i$.
 
It is worth stressing that, {while } the fit can always be applied {so that } some values of the {vortex } parameters {can be } obtained, the analytical inversion approach requires the strict respect of the condition \eqref{eq:domain} to extract the vortex parameter. 

\section{Experimental results and discussion}
\label{sec:experimental}
The measurements are performed at fixed, cryogenic temperature by putting the resonator inside an Oxford Instruments helium-flow CF-Dynamic Cryostat, placed between the polar expansions of a conventional electromagnet Bruker B-E25v. 
Sample thermalization is obtained by setting the cryostat temperature slightly lower than the desired temperature, through the helium flow and through heaters operated by an Oxford Instruments temperature controller ITC503S, and then by fine controlling resistive heaters mounted on the resonator.
The scattering coefficients of the resonator are measured by means of a R\&S ZVA-40 VNA, connected to the resonator through a $\sim1\,$m long cryogenic and nonmagnetic phosphor-bronze coaxial line plus a $\sim1.5\;$m long standard coaxial line at room temperature. All coaxial cables have 40 GHz cutoff frequencies and mount K connectors. 
The desired temperature is reached in zero-field cooling conditions, and then measurements are performed with a magnetic field $H$ applied perpendicularly to the sample surface. 
The sample used for the present discussion is a nominally $d=240\;$nm thick FeSe$_{0.5}$Te$_{0.5}$ superconductor film, with critical temperature ${T_c\sim18\;\rm{K}}$, deposited on a CaF$_2$ substrate.
Further details are reported in \cite{Braccini2013, Pompeo2020a}. 

A typical measurement at $T=10.00(5)\;$K is reported in Fig. \ref{fig:DZi}, where $\Delta Z_1$ and $\Delta Z_2$ for the two modes are reported in terms of real and imaginary parts vs the field $H$. 
The uncertainties are computed by taking ${u(Q_i)/Q_i=1\%}$ and ${u(\nu_{res,i})/\nu_{res,i}=2\times10^{-7}}$ as dominant uncertainties arising from the partial calibration of the {cryogenic} measurement line \cite{Torokhtii2020}.
The relative uncertainties on numerically computed $G_i$ can be typically in the $1\%-8\%$.  {With our estimates on the uncertainties of geometrical dimensions and permittivity, we estimate ${u(G_i)/G_i=3\%}$}. Such figure can improve with very accurate values for the permittivity \cite{Krupka1998, Krupka1999}. 
The relative uncertainty ${u(d)/d}$ depends on the technique used for its determination. The determination on the basis of the growth rate of the epitaxial film is typically $10\%$, while the use of atomic scanning force microscopes \cite{Bienias1998} yields $\sim1\%)$. Here we take ${u(d)/d}=5\%$.

In Fig. \ref{fig:vortexparameters} we report the vortex parameters extracted through the analytical and fit approaches discussed in the previous Section. 
It can be seen that in our case {$u(G_i)/G_i > u(\Delta r_i)/\Delta r_i,u(\Delta x_i)/\Delta x_i$}, {so that }the analytical inversion approach yields much smaller uncertainties on $\nu_c$ and $\chi$, while {on the other hand }the curve fit provides a smaller $u(\rhoff)$.
\footnote{A slight systematic difference between the values obtained with the different methods can be noted. Since this difference is well within the uncertainty bars ($k=1$), it could seem irrelevant. However, we verified that the two approaches always provide coincident values when tested with numerically simulated data. The slight difference obtained on real data can be considered a further test for the applicability of the model.
As an example, wideband measurements on the cuprate superconductor YBa$_2$Cu$_3$O$_{7-\delta}$ \cite{Wu1995} showed that below $\sim1-4\;$GHz, depending on the $T$ and $H$ ranges, the model of Eq. \eqref{eq:rhovmCC} does not apply. The little difference in the results of {Fig}.\ref{fig:vortexparameters} could be a precursor of a similar effect.} 
\begin{figure}[hbt!]
\centering
\includegraphics[width=0.8\columnwidth]{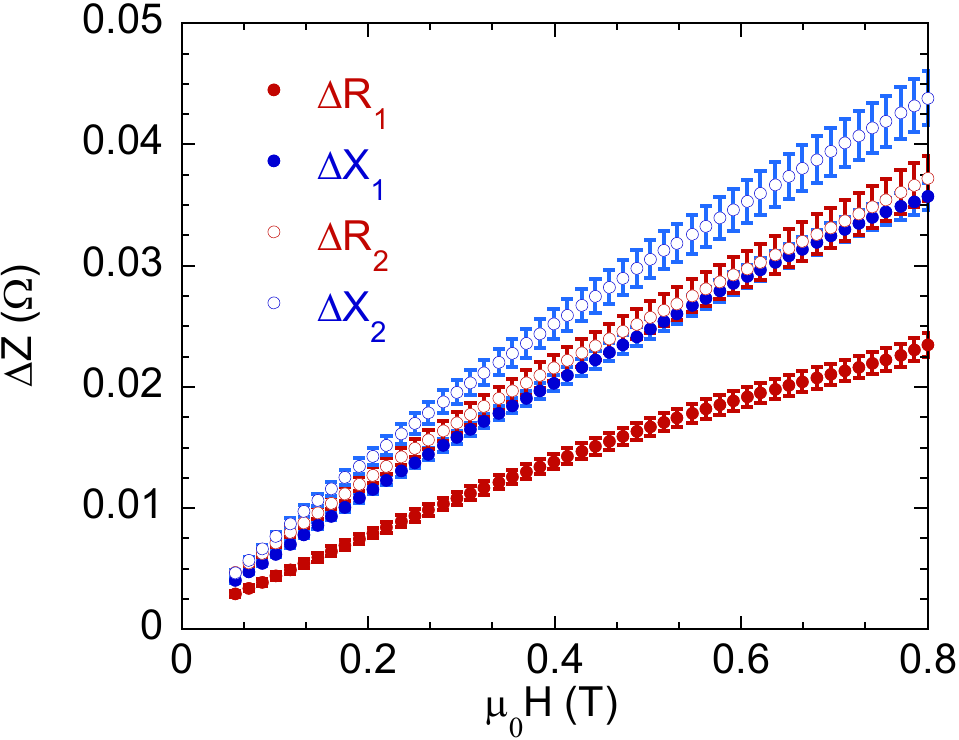}
\caption{$\Delta Z_i(H)$ at two frequencies, $\nu_{res,i}=\{16.4, 26.6\}\;$GHz, vs field $H$ at fixed \mbox{$T=10.00(5)\;$K}.}
\label{fig:DZi}
\end{figure}

The obtained results are very sound on physical grounds: $\nu_c$ and $\chi$ are nearly field independent, while $\rhoff$ is almost linear vs $H$, thus consistent with general models for the flux flow resistivity \cite{Bardeen1965}.
\begin{figure}[hbt!]
\centering
\includegraphics[width=0.8\columnwidth]{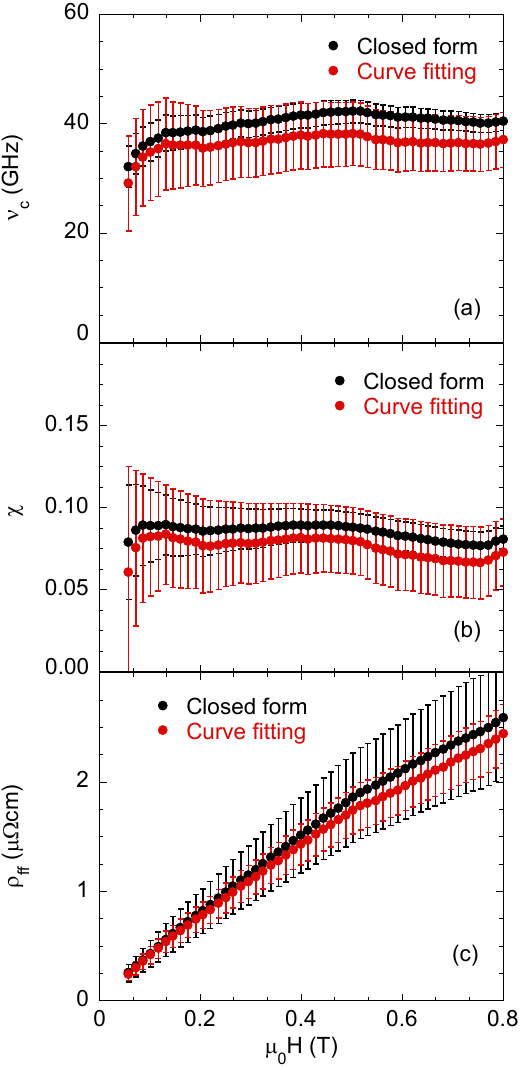}
\caption{{(a) Characteristic frequency $\nu_c$, (b) creep factor $\chi$ and (c) flux flow resistivity $\rhoff$ at $T=10.00(5)\;$K  extracted from the measurements of $\Delta Z$.}}
\label{fig:vortexparameters}
\end{figure}
The obtained $\nu_c$ is within a factor of 2.5 with respect to values reported in literature for $\nu_{c}(\chi=0)$ both in thin films \cite{PompeoEUCAS2019} and single crystals \cite{Okada2015}: considering the underestimation involved in the assumption $\chi=0$, as discussed in Sec. \ref{sec:intro}, together with the differences between distinct samples, this result is not unreasonable and it underlines the need of accurate measurements of the fundamental parameter $\nu_c$. 
The creep factor, although small, is not negligible even at this low temperature, as often naively assumed: this is consistent with results obtained with wide-band measurements in the same temperature range in other superconductors like Nb \cite{Silva2011}.

\subsection{Comparison with single frequency determinations}
It is of interest to compare the characteristic frequency value $\nu_c$, as obtained with the dual frequency ($\nu_1$ and $\nu_2$) method here described, to the value which would be obtained from a single frequency measurement, i.e. {$\nu_{p0}\coloneqq\nu_c(\chi=0)$ } from $\nu_1$ only (as an example). In the latter case, one obtains ${\nu_{p0}=25(1)\;{\rm GHz}}$: it can be seen that $\nu_{p0}$ is a heavy underestimation of $\nu_{c}$. Since measurements are often employed to compute values for $\rho_v$ or $Z$ at different frequencies, it is useful to stress that the significant underestimate inherent in $\nu_{p0}$ determines a much amplified underestimation on $\rho_{v}$ and hence $\Delta Z=\rho_{v}/d$ extrapolated to lower frequencies.
For example, at the selected field value ${\mu_0H=0.5\;}$T, using $\nu_{p0}$, 
{${\chi=0}$ and the corresponding flux flow resistivity $1.4(1)\times10^{-8}\;{\rm\Omega m}$, }
through Eq. \eqref{eq:rhovmCC} {$\Delta Z_{sim,0}=\rho_{v}/d$ } vs the frequency $\nu$ can be numerically computed {(here, the ``sim'' subscript highlights that the $Z_{sim}$ is numerically computed, while the ``0'' subscript highlights that the computation is done for zero creep)}. Analogously, the reconstructed frequency dependence of {$\Delta Z_{sim,\chi}$}, where the ``$\chi$'' subscript highlights that the experimentally obtained $\chi$ value is used, can be computed from the above reported values for $\nu_c$ and $\chi$, and the corresponding ${\rhoff=2.0(2)\times10^{-8}\;{\rm\Omega m}}$. The results are reported in Fig. \ref{fig:DZsim}a, together with the relative discrepancies
{$\varepsilon_R\coloneqq(R_{sim,\chi}-R_{sim,0})/R_{sim,\chi}$ and
 $\varepsilon_X\coloneqq(X_{sim,\chi}-X_{sim,0})/X_{sim,\chi}$ }
in Fig. \ref{fig:DZsim}b. 
\begin{figure}[hbt!]
\centering
\includegraphics[width=0.8\columnwidth]{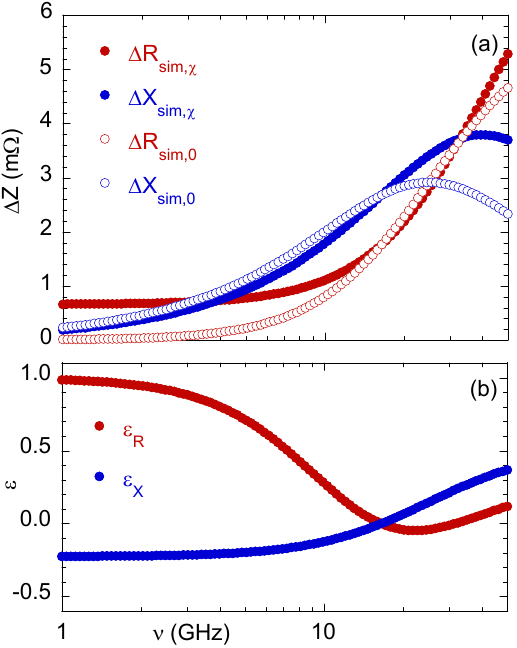}
\caption{Numerically computed {$\Delta Z_{sim}$ } -- through Eq. \eqref{eq:rhovmCC} --  at $\mu_0H=0.5\;{\rm T}$ extrapolating the results of a dual frequency measurement (subscript ``$\chi$'') and single frequency measurement, (subscript ``0''). Panel (a): {$\Delta Z_{sim}$ } vs frequency.
Panel (b): relative {discrepancies $\varepsilon_R$ and
$\varepsilon_X$, on $\Delta R_{sim}$ and $\Delta X_{sim}$ respectively, } vs frequency.
}
\label{fig:DZsim}
\end{figure}
It can be seen that, in particular, the relative discrepancy on $R$ can be as high as 100\%, thus making the single frequency measurement completely unreliable in estimating the material behaviour at lower frequencies.

\section{Summary}
\label{sec:conclusions}
We have addressed the problem of a reliable extraction of the main parameters governing the complex resistivity $\rhovm$ and then the surface impedance of superconducting films in a dc magnetic field. We have recalled the main model for the high-frequency response, underlining the relevant role of the characteristic frequency $\nu_c$ in determining the response of a superconductor in a magnetic field. In fact, $\nu_c$ is a synthetic measure of the superconductor performance (the higher, the better). 
We have described and experimentally illustrated a dual mode resonator that can be used to measure $\rhovm$ on a superconducting thin film at two distinct frequencies to correctly extract the relevant vortex parameters, and in particular $\nu_c$.
We have quantitatively shown that routinely performed single frequency measurements of $Z$
lead to heavy underestimations of $\nu_c$ and that the often neglected creep $\chi$ has to be taken into account, together with an accurate determination of $\nu_c$, to meaningfully extrapolate the $Z$ frequency dependence.
We have analysed the uncertainties resulting on the superconducting parameters, describing two alternative methods -- namely, an analytical, closed form, approach and a standard curve fitting -- to extract the mentioned quantities from the measurements. The advantages of both approaches have been highlighted, showing that the analytical one allows, in general, smaller uncertainties on $\nu_c$ and $\chi$. Within this discussion, a general examination of the uncertainties attainable by the method in function of the possible frequency ranges of $\nu_c$, and the subsequent criteria for its optimization, have been discussed.
The results here presented are of relevance to design and optimize cutting-edge experiments, such as cavities for dark matter research, or for impressively large infrastructure, such as the beam screen of the Future Circular Collider, and for fundamental research.

\section{Acknowledgments}

Work partially supported by MIUR-PRIN project \mbox{``HIBiSCUS''} - grant no. 201785KWLE.
The authors warmly thank Giulia Sylva and Valeria Braccini at CNR-SPIN (Genova, Italy) for providing the FeSeTe sample.

{
\appendix
\section{Appendix}
\label{sec:appendix}
The explicit solution of the system of equations \eqref{eq:system} is:
\begin{subequations}
\label{eq:solution}
\begin{align}
\label{eq:solution_nuc}
\frac{\nu_c}{\nu_1}&=\frac{(p^2-1)\r1\r2+\sqrt{\Lambda}}{2(p\r1-\r2)}\\ \nonumber
\Lambda&=\r1^2\r2^2(p^2-1)^2-4p(\r2-\r1 p)(p\r2-\r1)\\ 
\label{eq:solution_chi}
\chi&=\frac{\frac{\nu_c}{\nu_i}-\r{i}}{\frac{\nu_c}{\nu_i}\left(\frac{\nu_c}{\nu_i}\r{i}+1\right)}, i={1,2}
\end{align}
\end{subequations}
where $p=\nu_2/\nu_1>1$.
 In the above, Eq. \eqref{eq:solution_chi} shows that, once $\nu_c$ is determined, $\chi$ can be indifferently computed through the two reported equivalent expressions. {Moreover, the domain of the $\nu_c(\r1, \r2)$ function can be easily computed, yielding the result:
\begin{eqnarray}
\label{eq:domainapp}
\frac{\r1}{p}\leq\r2\leq p\r1 
\end{eqnarray}
}
{After Eq.s\eqref{eq:solution}, it is worth noting that the uncertainty on $\nu_c$,  $u(\nu_c)$:
\begin{eqnarray}
\label{eq:u_nuc}
\nonumber
u^2(\nu_c)=
\sum_{i=1}^2\left(\frac{\partial \nu_c}{\partial \r{i} }\right)^2\times\\
\times\left(\frac{1}{\Delta r_i}\right)^2\left(u^2(\Delta x_i)+\r{i}^2 u^2(\Delta r_i)\right)
\end{eqnarray}
can be evaluated} in closed form from the experimentally measured quantities $\r i$ and from the evaluation of the four uncertainties $u(\Delta r_i)$ and $u(\Delta x_i)$.
With the particularizations proposed in {Sec}. \ref{sec:optimal}, Eq. \eqref{eq:u_nuc} can be recast in the following:
\begin{eqnarray}
\label{eq:sensitivity_nu0}
\nonumber
u^2(\nu_c)=\left({u(\Delta r_1) G_1 d /\rhoff}\right)^2\times\\
\times\sum_{i=1}^2\left(\frac{\partial \nu_c}{\partial \r{i} }\right)^2(1+\r{i}^2)
\left(\frac{1}{\frac{\rho_{v1,i}}{\rhoff}}\right)^2 s_i^2
\end{eqnarray}
where for compactness' sake, $s_i$ has been introduced with $s_1=1$ and $s_2=s$.

The expressions solving the system Eq. \eqref{eq:system} for the flux flow resistivity are:
\begin{eqnarray}
\label{eq:rhoffi}
\rhoffn{i}=G_i d \Delta r_i \frac{1+\left(\frac{\nu_i}{\nu_c}\right)^2}{\chi+\left(\frac{\nu_i}{\nu_c}\right)^2}
\end{eqnarray}
}

\bibliographystyle{elsarticle-num}

\begin{thebibliography}{10}
\expandafter\ifx\csname url\endcsname\relax
  \def\url#1{\texttt{#1}}\fi
\expandafter\ifx\csname urlprefix\endcsname\relax\def\urlprefix{URL }\fi
\expandafter\ifx\csname href\endcsname\relax
  \def\href#1#2{#2} \def\path#1{#1}\fi

\bibitem{Posen2017}
S.~Posen, D.~L. Hall, {Nb$_3$Sn superconducting radiofrequency cavities:
  fabrication, results, properties, and prospects}, Supercond. Sci. Technol. 30
  (2017) 033004.

\bibitem{Cichalewski2020}
W.~Cichalewski, J.~Sekutowicz, A.~Napieralski, R.~Rybaniec, J.~Branlard,
  V.~Ayvazyan, {Continuous Wave Operation of Superconducting Accelerating
  Cavities With High Loaded Quality Factor}, IEEE Trans. Nucl. Sci. 67~(9)
  (2020) 2119--2127.

\bibitem{Calatroni2017b}
S.~Calatroni, R.~Vaglio, {Surface Resistance of Superconductors in the Presence of a DC Magnetic Field: Frequency and Field Intensity Limits}, IEEE Trans. Appl. Supercond. 27~(5) (2017) 3500506.

\bibitem{Abada2019}
A.~Abada, M.~Abbrescia, S.~S. AbdusSalam, I.~Abdyukhanov, J.~A. Fernandez,
  A.~Abramov, M.~Aburaia, A.~O. Acar, P.~R. Adzic, P.~Agrawal, J.~A.
  Aguilar-Saavedra, J.~J. Aguilera-Verdugo, M.~Aiba, I.~Aichinger, G.~Aielli,
  Etal, {FCC-hh: The Hadron Collider. Future Circular Collider Conceptual
  Design Report Volume 3}, Eur. Phys. J. Spec. Top. 228 (2019) 755--1107.

\bibitem{Alesini2019}
D.~Alesini, C.~Braggio, G.~Carugno, N.~Crescini, D.~D. Agostino, D.~D.
  Gioacchino, R.~D. Vora, P.~Falferi, S.~Gallo, U.~Gambardella, C.~Gatti,
  G.~Iannone, G.~Lamanna, C.~Ligi, A.~Lombardi, R.~Mezzena, A.~Ortolan,
  R.~Pengo, N.~Pompeo, A.~Rettaroli, G.~Ruoso, E.~Silva, C.~C. Speake,
  L.~Taffarello, S.~Tocci, {Galactic axions search with a superconducting
  resonant cavity}, Phys. Rev. D 99 (2019) 101101(R).

\bibitem{Alesini2020}
D.~Alesini, C.~Braggio, G.~Carugno, N.~Crescini, D.~D'Agostino, D.~{Di
  Gioacchino}, R.~{Di Vora}, P.~Falferi, U.~Gambardella, C.~Gatti, G.~Iannone,
  C.~Ligi, A.~Lombardi, G.~Maccarrone, A.~Ortolan, R.~Pengo, C.~Pira,
  A.~Rettaroli, G.~Ruoso, L.~Taffarello, S.~Tocci, {High quality factor
  photonic cavity for dark matter axion searches}, Rev. Sci. Instrum. 91~(9)
  (2020) 94701.

\bibitem{Collin1992}
R.~E. Collin, {Foundation for Microwave Engineering}, 2nd Edition, Singapore:
  McGraw-Hill International Editions, 1992.
  
{\bibitem{Landau1981}
L.~D. Landau and E.~M. Lifshifts, {Physical Kinetics}, 2nd Edition, Pergamon Press, 1981.

\bibitem{jackson1999book} J.~D.~Jackson, {Classical Electrodynamics}, 3rd Edition, John Wiley \& Sons Inc., USA, 1999}

\bibitem{PompeoImeko2017a}
N.~Pompeo, K.~Torokhtii, E.~Silva, {Surface impedance measurements in thin
  conducting films: Substrate and finite-thickness-induced uncertainties}, in:
  Proc. IEEE Int. Instrum. Meas. Technol. Conf. 22--25 May Turin Italy, no.~3,
  2017, pp. 1--5.

\bibitem{Tinkham1996book}
M.~Tinkham, {Introduction to Superconductivity}, 2nd Edition, McGraw-Hill,
  Inc., New York, NY, USA, 1996.

\bibitem{Golosovsky1996}
M.~Golosovsky, M.~Tsindlekht, D.~Davidov, {High-frequency vortex dynamics in
  YBa$_2$Cu$_3$O$_7$}, Supercond. Sci. Technol. 9~(1) (1996) 1--15.

\bibitem{PompeoLTP2020}
N.~Pompeo, A.~Alimenti, K.~Torokhtii, E.~Silva, {Physics of vortex motion by
  means of microwave surface impedance measurements (Review article)}, Low
  Temp. Phys. 46~(4) (2020) 343--347.

\bibitem{Pompeo2008}
N.~Pompeo, E.~Silva, {Reliable determination of vortex parameters from
  measurements of the microwave complex resistivity}, Phys. Rev. B 78~(9)
  (2008) 094503.

\bibitem{Gittleman1966}
J.~I. Gittleman, B.~Rosenblum, {Radio-frequency resistance in the mixed state
  for subcritical currents}, Phys. Rev. Lett. 16~(17) (1966) 734--736.

\bibitem{Song2009}
C.~Song, T.~Heitmann, M.~DeFeo, K.~Yu, R.~McDermott, M.~Neeley, J.~M. Martinis,
  B.~Plourde, {Microwave response of vortices in superconducting thin films of
  Re and Al}, Phys. Rev. B 79~(17) (2009) 174512.

\bibitem{Silva2011}
E.~Silva, N.~Pompeo, S.~Sarti, {Wideband microwave measurements in
  Nb/Pd$_{84}$Ni$_{16}$/Nb structures and comparison with thin Nb films},
  Supercond. Sci. Technol. 24~(2) (2011) 24018.

\bibitem{Silva2016}
E.~Silva, N.~Pompeo, K.~Torokhtii, S.~Sarti, {Wideband Surface Impedance
  Measurements in Superconducting Films}, IEEE Trans. Instrum. Meas. 65~(5)
  (2016) 1120--1129.

\bibitem{Booth1994}
J.~C. Booth, D.~H. Wu, S.~M. Anlage, {A broadband method for the measurement of
  the surface impedance of thin films at microwave frequencies}, Rev. Sci.
  Instrum. 65~(6) (1994) 2082--2090.

\bibitem{Kitano2008a}
H.~Kitano, T.~Ohashi, A.~Maeda, {Broadband method for precise microwave
  spectroscopy of superconducting thin films near the critical temperature.},
  Rev. Sci. Instrum. 79~(7) (2008) 074701.

\bibitem{Felger2013}
M.~M. Felger, M.~Dressel, M.~Scheffler, {Microwave resonances in dielectric
  samples probed in Corbino geometry: Simulation and experiment}, Rev. Sci.
  Instrum. 84 (2013) 114703.


\bibitem{Kobayashi1998}
Y.~Kobayashi, H.~Yoshikawa, {Microwave measurements of surface impedance of
  high-T$_c$ superconductors using two modes in a dielectric rod resonator},
  IEEE Trans. Microw. Theory Tech. 46~(12) (1998) 2524--2530.

\bibitem{Jacob2001}
M.~V. Jacob, J.~Mazierska, K.~Leong, J.~Krupka, {Simplified Method for
  Measurements and Calculations of Coupling Coefficients and Q o Factor of
  High-Temperature Superconducting Dielectric Resonators}, IEEE Trans. Microw.
  Theory Tech. 49~(12) (2001) 2401--2407.

\bibitem{Lue2002}
H.-t. Lue, J.-t. Lue, T.-y. Tseng, {Microwave Penetration Depth Measurement for High Tc Superconductors by Dielectric Resonators}, IEEE Trans. Instrum. Meas. 51~(3) (2002) 433--439.

\bibitem{Cherpak2003}
N.~Cherpak, S.~Member, A.~Barannik, Y.~Filipov, Y.~Prokopenko, S.~Vitusevich,
  {Accurate Microwave Technique of Surface Resistance Measurement of Large-Area
  HTS Films Using Sapphire Quasi-Optical Resonator}, IEEE Trans. Appl.
  Supercond. 13~(2) (2003) 3570--3573.

\bibitem{Hanaguri2003}
T.~Hanaguri, K.~Takaki, Y.~Tsuchiya, A.~Maeda, {An instrument for low- and
  variable-temperature millimeter-wave surface impedance measurements under
  magnetic fields}, Rev. Sci. Instrum. 74~(10) (2003) 4436--4441.

{\bibitem{Hashimoto2003} T.~Hashimoto and Y.~Kobayashi, Frequency Dependence Measurements of Surface Resistance of Superconductors Using Four Modes in a Sapphire Rod Resonator, IEICE Trans. Electron. E86-C~(8) (2003) 1721--1728.}
  
\bibitem{Cherpak2005}
N.~Cherpak, A.~Barannik, S.~Bunyaev, Y.~Prokopenko, S.~Vitusevich,
  {Measurements of Millimeter-Wave Surface Resistance and Temperature
  Dependence of Reactance of Thin HTS Films Using Quasi-Optical Dielectric
  Resonator}, IEEE Trans. Appl. Supercond. 15~(2) (2005) 2919--2922.

  
\bibitem{Ghigo2005a}
G.~Ghigo, D.~Andreone, D.~Botta, A.~Chiodoni, R.~Gerbaldo, L.~Gozzelino,
  F.~Laviano, B.~Minetti, E.~Mezzetti, {Non-uniform columnar defect
  implantation in YBCO coplanar resonators for the control of vortex-induced
  microwave dissipation and nonlinearity}, Supercond. Sci. Technol. 18 (2005)
  193--199.

\bibitem{Huttema2006}
W.~A. Huttema, B.~Morgan, P.~J. Turner, W.~N. Hardy, X.~Zhou, D.~A. Bonn,
  R.~Liang, D.~M. Broun, {Apparatus for high-resolution microwave spectroscopy
  in strong magnetic fields}, Rev. Sci. Instrum. 77~(2) (2006) 023901.

 
\bibitem{Yeh2013}
J.-h. Yeh, S.~M. Anlage, {In situ broadband cryogenic calibration for two-port
  superconducting microwave resonators}, Rev. Sci. Instrum. 84 (2013) 034706
  1--8.


\bibitem{Pompeo2014}
N.~Pompeo, K.~Torokhtii, E.~Silva, {Dielectric Resonators for the Measurements
  of the Surface Impedance of Superconducting Films}, Meas. Sci. Rev. 14~(3)
  (2014) 164--170.

\bibitem{Ohshima2018}
S.~Ohshima, N.~Takanashi, A.~Saito, K.~Nakajima, T.~Nagayama, {Effect of Si-Ion
  Irradiation on Microwave Surface Resistance in YBa2Cu3Oy Thin Films in
  Magnetic Fields}, IEEE Trans. Appl. Supercond. 28~(4) (2018) 1--4.

\bibitem{Alimenti2019a}
A.~Alimenti, K.~Torokhtii, E.~Silva, N.~Pompeo, {Challenging microwave resonant measurement techniques for conducting material characterization}, Meas. Sci. Technol. 30 (2019) 065601.

{\bibitem{iec}
International Electrotechnical Commission, {IEC 61788-7:2020 Superconductivity - Part 7: Electronic characteristic measurements - Surface resistance of high-temperature superconductors at microwave frequencies} (2020).
}

  
\bibitem{Powell1998}
J.~Powell, A.~Porch, R.~Humphreys, F.~Wellh\"{o}fer, M.~Lancaster, C.~Gough,
  {Field, temperature, and frequency dependence of the surface impedance of
  YBa$_2$Cu$_3$O$_7$ thin films}, Phys. Rev. B 57~(9) (1998) 5474--5484.

{
\bibitem{Hakki1960} B.~W.~Hakki and P.~D.~Coleman, A Dielectric Resonator Method of Measuring Inductive Capacities in the Millimeter Range, IRE Trans. Microw. Theory Tech. 8~(4) (1960) 402--410.}
  
\bibitem{PompeoIMEKO2020}
N.~Pompeo, K.~Torokhtii, A.~Alimenti, E.~Silva, {Dual frequency resonator for
  the correct determination of the in-field surface impedance frequency
  dependence of superconductors}, in: 24th IMEKO TC4 Int. Symp., 2020, pp.
  75--79.
  
{
\bibitem{Pompeo2017}
N.~Pompeo, K.~Torokhtii, and E.~Silva, {Design and test of a microwave resonator for the measurement of resistivity anisotropy}, Measurement 98 (2017), 414--420.
}

\bibitem{Pompeo2017a}
N.~Pompeo, K.~Torokhtii, F.~Leccese, A.~Scorza, S.~Sciuto, E.~Silva, {Fitting
  strategy of resonance curves from microwave resonators with non-idealities},
  in: Proc. IEEE Int. Instrum. Meas. Technol. Conf. 22--25 May Turin Italy,
  Vol.~21, 2017, pp. 1--6.

\bibitem{Torokhtii2019a}
K.~Torokhtii, A.~Alimenti, N.~Pompeo, E.~Silva, {Uncertainty in uncalibrated
  microwave resonant measurements}, in: 24th IMEKO TC4 Int. Symp., 2019, pp.
  1--5.

\bibitem{Staelin1994book}
D.~H. Staelin, A.~W. Morgenthaler, J.~A. Kong, {Electromagnetic Waves},
  Prentice-Hall, Inc., 1994.

\bibitem{Zic2016}
M.~\v{Z}ic, {An alternative approach to solve complex nonlinear least-squares
  problems}, J. Electroanal. Chem. 760 (2016) 85--96.

\bibitem{More1980}
J.~J. More, B.~S. Garbow, K.~E. Hillstrom, {User guide for MINPACK-1. [In
  FORTRAN]}, Tech. rep., United States (1980).

\bibitem{Braccini2013}
V.~Braccini, S.~Kawale, E.~Reich, E.~Bellingeri, L.~Pellegrino, A.~Sala,
  M.~Putti, K.~Higashikawa, T.~Kiss, B.~Holzapfel, C.~Ferdeghini, {Highly
  effective and isotropic pinning in epitaxial Fe(Se,Te) thin films grown on
  CaF$_2$ substrates}, Appl. Phys. Lett. 103 (2013) 172601.

\bibitem{Pompeo2020a}
N.~Pompeo, K.~Torokhtii, A.~Alimenti, G.~Sylva, V.~Braccini, E.~Silva, {Pinning
  properties of FeSeTe thin film through multifrequency measurements of the
  surface impedance}, Supercond. Sci. Technol. 33~(11) (2020) 114006.

\bibitem{Torokhtii2020}
K.~Torokhtii, A.~Alimenti, F.~Rizzo, A.~Augieri, G.~Celentano, A.~Frolova,
  E.~Silva, N.~Pompeo, {High frequency vortex dynamics in
  {YBa}$_2$Cu$_3$O$_{7-x}$ with Ba$_2$YTaO$_6$-Ba2YNbO$_6$ nanodefects}, J.
  Phys. Conf. Ser. 1559 (2020) 12043.

\bibitem{Krupka1998}
J.~Krupka, K.~Derzakowski, B.~Riddle, J.~Baker-Jarvis, {A dielectric resonator
  for measurements of complex permittivity of low loss dielectric materials as
  a function of temperature}, Meas. Sci. Technol. 9~(10) (1998) 1751--1756.

\bibitem{Krupka1999}
J.~Krupka, K.~Derzakowski, M.~Tobar, J.~Hartnett, R.~G. Geyer, {Complex
  permittivity of some ultralow loss dielectric crystals at cryogenic
  temperatures}, Meas. Sci. Technol. 10~(5) (1999) 387--392.

\bibitem{Bienias1998}
M.~Bienias, S.~Gao, K.~Hasche, R.~Seemann, K.~Thiele, {A metrological scanning
  force microscope used for coating thickness and other topographical
  measurements}, Appl. Phys. A 66~(1) (1998) S837--S842.

\bibitem{Wu1995}
D.-H. Wu, J.~Booth, S.~Anlage, {Frequency and field Variation of Vortex
  Dynamics in YBa$_2$Cu$_3$O$_{7-\delta}$}, Phys. Rev. Lett. 75~(3) (1995)
  525--528.

\bibitem{Bardeen1965}
J.~Bardeen, M.~Stephen, {Theory of the Motion of Vortices in Superconductors},
  Phys. Rev. 140~(4A) (1965) 1197--1207.

\bibitem{PompeoEUCAS2019}
N.~Pompeo, A.~Alimenti, K.~Torokhtii, G.~Sylva, V.~Braccini, E.~Silva,
  {Microwave properties of Fe(Se,Te) thin films in a magnetic field: pinning
  and flux flow}, J. Phys. Conf. Ser. 1559 (2020) 012055.

\bibitem{Okada2015}
T.~Okada, F.~Nabeshima, H.~Takahashi, Y.~Imai, A.~Maeda, {Exceptional
  Suppression of Flux-Flow Resistivity in FeSe$_{0.4}$Te$_{0.6}$ by Back-Flow
  from Excess Fe Atoms and Se/Te Substitutions}, Phys. Rev. B 91 (2015) 054510.

\end{thebibliography}

\end{document}